# Langues en danger et multilinguisme numérique

**Mokhtar BEN HENDA**
*MCF-HDR en Sciences de l'Information et de la Communication,
MICA EA-4426, Université Bordeaux Montaigne
mbenhenda@u-bordeaux3.fr*

RÉSUMÉ : À l'ère de la mondialisation et des réseaux numériques, les langues dites « minorées » ou « en danger » sont devant un double dilemme : réussir leur modernité numérique en acceptant un « douloureux » réaménagement linguistique ou aller vers une extinction progressive face à des langues hégémoniques « prédatrices » qui dominent l'espace digital sur les réseaux.
Langues orales et écritures en danger non romanisées sont les plus concernées par les mesures de protection de la diversité culturelle et linguistique sur Internet. L'audiovisuel numérique et l'encodage multi-écriture par Unicode leur offrent des alternatives innovantes, consensuelles et normalisées, pour s'affirmer. Tout dépendra de la synergie que leurs communautés de pratiques créeront autour d'elles pour les placer au cœur du débat sur la fracture numérique.

MOTS-CLÉS : *Multilinguisme numérique, diversité linguistique, langues en danger, encodage numérique, audiovisuel numérique, aménagement linguistique.*

ABSTRACT: *In the era of globalization and the digital networks, the so-called "minored" or "endangered" languages are facing a twofold dilemma: either succeed their digital modernity by accepting a "painful" linguistic management, or slide towards a slow extinction in front of hegemonic and "predatory" languages which dominate the digital networks.*
*Oral languages and minored not-Romanized writings are the most concerned by the protective measures of the cultural and linguistic diversity on the Internet. The digital broadcasting and the Unicode multi-writing encoding system are providing them with innovative, consensual and standardized alternatives to survive. Then, it depends on the synergy that their communities of practice will generate to place them at the heart of the debate on the digital divide.*

La disparition des langues et des écritures est un fait auquel l'Humanité a toujours été confrontée. Les langues et les écritures sont des entités vivantes, en constante évolution, et leur survie ou extinction est tributaire de facteurs complexes et variés. Aujourd'hui, au-delà des inégalités entre langues « dominantes » et langues dominées », c'est plutôt le rythme auquel des langues et des écritures disparaissent qui est inquiétant.

Avec les nouvelles technologies numériques, un nouvel espoir est né de voir l'Internet servir d'environnement palliatif au déclin de beaucoup de langues et d'écritures. Avec des moyens et des méthodes technologiques de traitement linguistique plus souples, puissants et faciles d'accès, un réaménagement salvateur des langues en danger par le numérique est-il possible ?

**Délimitation d'un périmètre de « danger linguistique »**

Le sujet des langues menacées sur Internet est un problème aux racines très anciennes lié au caractère dynamique des langues humaines qui évoluent, mutent, se transforment, meurent et renaissent constamment. D'où le problème persistant de savoir comment et selon quels critères identifier une langue est en danger ou oubliée. Les études ont toujours été confrontées à la complexité du phénomène due à son caractère hétéroclite et transversal.

Plusieurs tentatives de classification des facteurs de risque pour une langue ont toutefois permis d'en esquisser les grandes lignes qui convergent vers des critères liés à la réduction du nombre de locuteurs, à l'incapacité d'assurer une communication intergénérationnelle, à la non-adaptabilité aux innovations, à l'éducation et à la production d'un patrimoine culturel consistant. L'Unesco établit dans ce sens une classification à cinq échelles : 1) une langue « sécurisée » est une langue largement parlée ; 2) une langue « vulnérable » n'est pas parlée par les enfants en dehors des foyers ; 3) une langue « en danger » n'est pas du tout parlée par les enfants ; 4) une langue « sérieusement en danger » n'est parlée que par des anciens ; 5) une langue dans un « état critique » est parlée occasionnellement par un petit nombre d'anciens[1].

Des propositions émanant des domaines de la linguistique, de la sociolinguiste et de la linguistique computationnelle, concordent aussi à classer les langues en trois catégories. Il y a d'abord des langues de travail ou « langues dominantes » qui sont largement utilisées pour la production des savoirs et la communication de masse. Il s'agit ensuite de langues de laboratoire ou « langues objets » qui sont considérées plus comme un support d'étude linguistique ou anthropologique qu'un moyen de communication ; et enfin des « langues oubliées » ou « exclues » qui ne servent dans aucun des deux cas précédents.

Dans les limites de ce papier, nous construisons notre acception de l'oubli des langues et de leur mise en danger sur la base de deux paramètres principaux. Le premier est socio-culturel et anthropologique : une langue serait considérée comme oubliée pour des raisons inhérentes d'abord à la réduction de son champ d'usage – donc à son nombre limité de locuteurs – puis au manque ou l'absence de son potentiel à pouvoir représenter un patrimoine culturel consistant, ensuite à sa non reconnaissance comme langue de communication dans un contexte national ou international. Il s'agit en fait d'une série de facteurs socio-culturels qui peuvent s'amplifier sous l'effet de facteurs politiques et économiques contraignants qui conduisent une langue à devenir une langue « oubliée » ou – à des degré moindres – une langue « exclue », « minorée » ou encore, comme dans la taxonomie de l'Unesco, une langue « en danger ».

Le deuxième paramètre qui nous intéresse plus particulièrement est de nature technique. Il s'agit de l'absence d'un cadre technologique approprié et suffisamment outillé, notamment pour des langues exclusivement orales, qui peinent encore à s'intégrer convenablement dans la dynamique mondiale du multilinguisme sur les réseaux. À l'ère d'Internet, le sort des langues orales, encore plus que celui des langues écrites, est au cœur du problème de la fragilité multilingue internationale. Étant beaucoup plus nombreuses que les langues écrites, les langues orales soulèvent des questions liées, non seulement à l'inégalité des langues dans le cyberespace, mais aussi à l'avenir de la diversité linguistique numérique mondiale.

---

[1] UNESCO AD HOC EXPERT GROUP, *Language Vitality and Endangerment*, Paris, Unesco, p. 17.

**Indicateurs socio-culturels de la fragilité linguistique mondiale**

Le diagnostic alarmant des langues « en danger » n'est pas une nouveauté. Claude Hagège l'a signalé depuis plus d'une décennie dans son passionnant plaidoyer *Halte à la mort des langues* : « comme les civilisations, les langues sont mortelles, et le gouffre de l'histoire est assez grand pour toutes »[2].

Des indicateurs factuels plus récents corroborent le bien-fondé de ce diagnostic. Les récentes études coordonnées par le Réseau mondial pour la diversité linguistique Maaya et publiées dans le rapport Net.Lang[3], sont très claires. Daniel Prado, Secrétaire exécutif de Maaya, élabore un bilan argumenté de statistiques très significatives comme l'estimation du linguiste britannique David Crystal selon laquelle « 50 % des langues de la planète seraient aujourd'hui parlées par moins de dix mille habitants » ou encore celle du sociolinguiste québécois Jacques Leclerc selon laquelle « les 74 premières langues de l'humanité sont parlées par 94 % des habitants de la planète »[4].

Dans son rapport annuel 2011, le *Summer Institute of Linguistics* (SIL) rapporte aussi que seulement entre 300 et 500 langues ont une présence effective dans le monde (mesurée en nombre de locuteurs et en patrimoine culturel). Ce nombre représenterait environ 6 % des quelque 6000 langues parlées, ce qui constituerait un indicateur significatif de l'état de la fragilité qui caractérise les autres 94 %. Selon Claude Hagège, 25 langues meurent chaque année[5] et environ tous les 15 jours, le dernier locuteur d'une langue s'éteint : « une érosion linguistique » qui aboutirait vers la fin de ce

---

[2] HAGÈGE (Claude), *Halte à la mort des langues*, Paris, Odile Jacob, 2000, p. 10.

[3] VANNINI (Laurent) ; LE CROSNIER (Hervé), dir., *Net.Lang : réussir le cyberespace multilingue*, Caen, C&F éditions, 2012.

[4] PRADO (Daniel), « Présence des langues dans le monde réel et le cyberespace », in VANNINI (Laurent) ; LE CROSNIER (Hervé), dir., *Net.Lang…, op. cit.*

[5] HAGÈGE (Claude), *op. cit.*, p. 404.

siècle à l'extinction d'environ un tiers à la moitié des langues parlées actuellement[6].

Dans le contexte numérique, la situation n'est pas plus brillante puisqu'à peine 5 % des langues du monde (350 environ) ont une présence réelle dans le cyberespace (locuteurs et contenus). Même au sein de ce petit nombre, il existe des écarts considérables. En 2013, Wikipédia aurait été dans 284 langues différentes largement dominées par l'anglais, l'allemand, le français et le néerlandais, et avec plus de 25 millions d'articles accessibles par environ 40 millions d'utilisateurs enregistrés. Word propose 126 langues dans son correcteur orthographique et Google se décline dans 145 langues (moins que la moitié des langues sur Internet). Google Livres contient environ 7 millions de documents dans uniquement 44 langues[7]. Parmi les 2000 langues d'Afrique (environ le tiers des langues du monde), seules 400 sont encodées dans Unicode.

**Vers une modernité linguistique numérique**

On peut affirmer aujourd'hui, qu'à l'heure de la société connectée, une langue qui n'est pas « accessible » (dans le sens instrumental) par ordinateur est une langue « menacée », une langue « peu ou mal dotée » voir « exclue ». Cela s'applique encore à des langues écrites dont l'internationalisation dépend de leur capacité d'afficher, transmettre, stocker et récupérer de façon fiable, reconnue, des contenus numériques dans leurs orthographes d'origine. Mais le problème se poserait avec plus d'intensité pour des langues seulement orales « pour lesquelles l'espace numérique représente un handicap fatal sauf à réaliser l'effort d'inventer une forme écrite et codifiable »[8]. Une majorité de langues orales dans le monde se voient menacées d'exclusion, voire d'extinction, juste parce qu'elles ne sont pas équipées d'un mécanisme d'encodage numérique qui assurerait leur présence sur les réseaux.

---

[6] EVANS (Nicholas), *Dying words: Endangered languages and what they have to tell us.* Wiley-Blackwell, 2013.

[7] PRADO (Daniel), *op. cit.*, p. 41.

[8] PAOLILLO (John) ; PIMIENTA (Daniel) ; PRADO (Daniel), *Mesurer la diversité linguistique sur Internet*, Paris, Unesco, 2005, p. 19.

Cette réalité est reconnue par le *World Language Documentation Centre* (WLDC)[9], organe de l'Unesco chargé des questions terminologiques : « Dans les pays développés, le web semble une réalité mondiale, mais il ne faut pas oublier qu'une grande partie des langues parlées dans le monde sont très peu représentées sur Internet [et que] dans un certain nombre de cas, cette situation est due à l'absence de systèmes de représentation des langues et de leurs variantes »[10]. Daniel Prado l'entérine aussi dans Net.Lang : « entre 90 et 95 % des langues de la planète n'ont pas d'écriture »[11]. Faudrait-il pour autant que les 95% des langues du monde se dotent d'un système d'écriture pour s'affranchir de l'exclusion numérique ? Une langue orale a-t-elle réellement besoin d'être écrite pour avoir sa place dans le monde numérique ? En d'autres termes, le lien entre langue orale et système d'écriture est-il vital dans l'univers numérique ?

Des enjeux profonds se jouent en effet autour de ces questions dans un débat au centre duquel se posent fortement des questions liées à l'encodage numérique des langues et des systèmes d'écriture et à la nécessité ou non du passage d'une langue orale à un stade écrit pour assurer sa survie et sa visibilité numérique.

**La transcription graphique : voie de salut pour les langues en danger ?**

Plusieurs solutions ont été trouvées pour doter les langues orales de systèmes de transcription graphique. Certaines sont sous la forme d'une translittération (transposition des phonèmes d'une langue orale en graphèmes d'un système d'écriture emprunté) comme le wolof qui peut être transcrit par des caractères arabes et latins. D'autres langues, pratiquées un peu partout en Afrique, aux Amériques, en Asie ou en Australie, ont été transcrites – assez tardivement (fin XIX$^e$ - début XX$^e$ siècle) – par des représentations graphiques héritées des traditions millénaires. Ces alphabets autochtones étaient, à l'origine, le résultat d'initiatives individuelles ou de consensus communautaires improvisés. Ce facteur a beaucoup ralenti leur visibilité pour avoir été élaborés en dehors d'un cadre linguistique et normatif structurant comme l'Alphabet Phonétique International. À l'opposé, la romanisation planifiée du « mandingue » et du « peul » a fait de ces langues les mieux connues aujourd'hui en Afrique occidentale.

---

[9] Chargé des technologies des langues, linguistique, normalisation terminologique et localisation, le WLDCT a été officiellement lancé le 9 mai 2007 au Siège de l'UNESCO à Paris. Il est dirigé par Christian Galinsky, secrétaire de l'ISO TC37, comité de terminologie de l'Organisation internationale de normalisation.

[10] UNESCO, *Lancement officiel du centre mondial de documentation sur les langues*, Unesco, Communication et information [En ligne], 2007, [http://www.unesco.org/new/fr/communication-and-information/] (11 janvier 2014).

[11] PRADO (Daniel), *op. cit.*, p. 38.

Des langues asiatiques et océaniennes, composées dans leur grande majorité de langues orales, ont également été translitérées par les missionnaires à partir du XIX$^e$ siècle par des conventions d'écriture et des systèmes de transcription alphabétiques adaptés, basés sur la phonologie des langues concernées. Le chinois, par exemple, dispose de plusieurs systèmes pour la translittération du mandarin en caractères latins et la Corée du Sud a développé en 2000 une norme de romanisation révisée pour la translittération du coréen.

Partout ailleurs, des familles de langues disposent de plusieurs systèmes d'écriture. Les langues berbères (kabyle, rifain, chleu, les nombreuses langues touareg…) et leurs nombreuses variantes parlées, sont transcrites en caractères arabes, latins et tifinagh. Les dialectes kurdes (sorani, kurmandji et zazaki) sont aussi transcrits en caractères latins, cyrilliques, arabes ou persans. Ces langues et tant d'autres ont dû converger vers des systèmes d'écriture uniforme avant d'être reprises dans un système d'encodage numérique. Il s'agit d'une forme d'aménagement linguistique dont nous parlerons plus loin comme voie stratégique cruciale pour survivre et être plus visible sur la scène internationale.

**L'encodage numérique : atouts et insuffisances pour les langues en danger**

En passant au stade numérique, beaucoup de langues et écriture indigènes – souvent de nature logographique ou syllabique – ont été confrontées au problème que leurs systèmes d'écriture contiennent beaucoup plus de caractères que ne le supporte un clavier d'ordinateur « standard ». Deux solutions ont souvent été prises en compte pour contourner cette contrainte. La première est la romanisation qui se traduit par un passage d'un système d'écriture autochtone à l'usage de l'alphabet romain. La deuxième est le développement de normes d'encodage numérique et de méthodes de saisie par clavier. Ces solutions ont été développées à une époque marquée par l'émergence de la notion de jeux de caractères codés qui avait permis la conception des systèmes de cartes perforées depuis la fin du XIX$^e$ siècle, puis de tables de codes informatiques utilisées par les ordinateurs depuis les années 1930. Les machines à encoder en mode numérique étaient cependant limitées à un usage réservé aux scientifiques et chercheurs dans des laboratoires. Elles fonctionnaient à base de systèmes d'encodage très variés et souvent incompatibles.

Quand le code ASCII (*American Standard Codification of information Interchange*) s'est imposé au début des années 1970 comme solution normative internationale, il encodait uniquement 128 caractères propres à la langue anglaise et à quelques autres langues romanisées. On était encore loin des soucis de l'équilibre linguistique dans le monde numérique. Mais cette primauté historique de l'ASCII explique en partie les raisons pour lesquelles l'anglais (ou disons les caractères latins non accentués) dominent encore le contexte numérique mondial. Plus tard, quand l'informatique et la téléinformatique sont sortis des laboratoires pour être accessibles à des usages sociaux grâce aux ordinateurs personnels et portables, le monde s'est trouvé face à une forme d'exclusion linguistique qui s'étendait au sein même de la famille de langues latines puisque les diacritiques (comme les accents en français) rendaient très compliqué l'usage correct et transparent de ces langues par les logiciels et les ordinateurs. Pour y remédier, un grand nombre de solutions d'encodages multilingues a été proposé par des chercheurs, des industriels et des institutions de standardisation nationales et internationales. Mais la majorité de ces solutions étaient incompatibles entre elles aggravant davantage le problème de la diversité linguistique numérique.

Vers la fin des années 1980, l'Organisation internationale de normalisation (ISO) a proposé une série de normes connues par leur nom de code « ISO/IEC 8859 » pour encoder une quinzaine de langues majoritairement européennes. C'était un bon début et un tournant important dans l'histoire de l'informatique multilingue, même s'il concernait un nombre limité de langues comparé au très large patrimoine mondial de langues et d'écritures. Cette famille de normes a eu l'avantage d'avoir préparé l'arrivée en 1991 du standard Unicode comme un moyen d'encoder un répertoire plus étendu de caractères compatibles avec tous les alphabets du monde. Un nouvel espoir est alors né pour les langues et les écritures minoritaires qui n'avaient ni les moyens ni les prédispositions nécessaires pour être inscrites assez tôt dans la course de l'informatique multilingue.

Unicode est un standard de codage universel indépendant des plates-formes et des systèmes d'exploitation des machines. Il fournit un moyen de coder tous les caractères existants dans « toutes » les langues et les écritures du monde ancien et moderne. Son avantage est qu'il ne fonctionne pas sur le principe de l'identification des langues mais plutôt selon une codification unique des caractères en totale abstraction de leurs formes (glyphes) et de leurs appartenances linguistiques. Cette démarche est innovante et économique en ressources du moment qu'il est possible d'encoder plusieurs langues en utilisant les mêmes caractères sans besoin de les dédoubler. Pour cela, Unicode est un système d'encodage particulièrement bien adapté au réseau Internet, car la nature mondiale du réseau exige des solutions qui fonctionnent dans n'importe quelle langue et n'importe quelle écriture.

Dans sa dernière version, Unicode couvre une centaine de systèmes d'écriture et une quinzaine de collections de symboles dans une répartition capable d'accueillir jusqu'à 1 114 112 caractères. Cette grande capacité est plus que suffisante pour tous les besoins de codage de caractères connus, y compris une couverture complète de tous les groupes minoritaires et les scripts historiques du monde[12]. L'inégale présence des langues et des écritures sur Internet ne peut donc être exclusivement attribuée à un problème d'espace dans Unicode, mais plutôt à des facteurs qui lui sont externes. Adapter un logiciel, un navigateur ou un ordinateur au standard Unicode n'est qu'une étape dans le processus d'internationalisation. Il faut fournir beaucoup plus de codes supplémentaires qui s'adaptent à des préférences culturelles ou des règles linguistiques comme la césure des mots, la rupture

---

[12] THE UNICODE CONSORTIUM, *The Unicode Standard Version 6.0 – Core Specification*, Mountain View, CA, The Unicode Consortium, 2011, p.1

des lignes, la bidirectionnalité des écritures ou la composition des caractères complexes (diacritiques). La complexité de ces opérations de nature linguistique et plutôt tributaire de logiciels et de protocoles de niveau supérieur. Ce détail est un autre facteur d'inégalité (voire de risque) pour beaucoup de langues mal équipées en produits d'ingénierie linguistique et de traitement automatisé des langues. La présence et la visibilité des langues sur Internet nécessitent incontestablement l'appui d'une ingénierie de recherche et d'une industrie linguistique. Beaucoup de langues « minoritaires », y compris des langues latines, n'ont pas atteint un seuil avéré de visibilité car il leur manque des produits d'appui issus de la recherche linguistique, technologique et industrielle.

En tant que consortium d'industriels, Unicode tient compte de ce principe de « rentabilité » pour accepter l'intégration ou le rejet de caractères ou de jeux entiers de caractères. Plusieurs propositions sont toujours en discussion ou rejetés[13]. Des écritures archaïques ou « mortes » dont les chercheurs n'ont pas encore exprimé l'intérêt de les numériser, ne sont pas non plus prises en compte. Pourtant, ce sont des écritures et des langues de valeurs anthropologiques et historiques confirmées comme les hiéroglyphes anatoliens des Hittites, l'albanais caucasien, le nabatéen, le pahlavi, etc.

L'intégration d'une écriture dans Unicode est une initiative volontariste et ascendante. C'est aux parties concernées de faire des propositions au consortium pour qu'il intègre leur système d'écriture. Or, tous les acteurs concernés par les questions de la numérisation des langues (communautés de locuteurs, institutions de recherche, États, consortiums…) n'ont pas la même prédisposition à le faire. On comprend alors pourquoi des écritures sont mal représentées – ou ne le sont du tout – dans ce codage universel.

**L'accès aux contenus en ligne dans les langues maternelles**

---

[13] Parmi les écritures actuelles non encore retenues : le « bassa vah », le « loma », et le « mendé kikakui », des écritures pour des langues orales du Liberia, Sierra Leone et sud-est de la Guinée, le « ganta » et le « modi » des écritures pour transcrire des dialectes indiens.

L'accès aux ressources numériques en ligne est un autre problème contraignant pour la diversité linguistique et la visibilité des langues sur Internet. Toutes les ressource et tous les services hébergés sur des serveurs en réseau nécessitent sans exception des adresses d'accès (aussi bien des URLs: *Uniform Resouce Locator* pour les produits du Web que des adresses de messagerie électronique). Ces dernières ont la caractéristique d'avoir toujours été – notamment dans la « racine » ou « domaine de tête » – exclusivement en caractères latins non accentués. Or, le nombre de locuteurs latins utilisant des caractères diacritiques ou des locuteurs monolingues non latins est très important et impacte lourdement le niveau d'usage et d'accessibilité à ces services et ressources sur les réseaux. Selon des statistiques rapportées par l'*Internet Names Authorization & Information Center* (INAIC), 500 millions de personnes dans le monde sont en ligne chaque jour dont seulement un tiers parle anglais comme langue maternelle. Même si les deux autres tiers savent déchiffrer les 26 lettres latines pour saisir l'adresse d'une ressource ou d'une personne, des catégories entières de la population mondiale en sont privées. Un érudit monolingue chinois ou un paysan mongol qui ne parle que sa langue maternelle, ne peut utiliser un ordinateur pour accéder à des contenus en ligne tant que ces contenus ont des adresses d'accès uniquement en caractères latins. Ce principe hérité d'un encodage ASCII hégémonique, contredit toute logique de réaliser un Internet multilingue.

Des solutions techniques ont commencé à être étudiées depuis 1996 jusqu'à ce que l'*Internet Engineering Task Force* (IETF) mette en place en 2008 un standard d'internationalisation des noms de domaines (IDN). L'idée principale derrière les IDN est d'utiliser – par un processus standardisé – des caractères non-ASCII dans la composition des adresses électroniques sur les réseaux. En 2009, l'*Internet Corporation for Assigned Names and Numbers* (Icann), l'organisme américain qui gère les noms de domaine Internet à l'échelle mondiale, accepte d'intégrer les IDN dans les noms des domaines de premier niveau (*Top Level Domain*). En 2010, la première adresse Internet en caractères non latins marque le commencement des modes d'accès multilingues sur le réseau.

La recherche dans ce domaine est encore récente, complexe et engage plusieurs acteurs ; ce qui l'expose aux enjeux de sphères d'influence et de « nombreuses luttes politiciennes »[14]. Mais l'avancement des travaux sur ce terrain laisse entrevoir que les noms de domaines multilingues progressent lentement – mais aussi concrètement – pour assurer dans l'avenir une meilleure démocratisation de l'accès à des sites web dans les langues natives des utilisateurs. Ils seraient ainsi un facteur important dans la transformation de l'Internet en une ressource véritablement mondiale et multilingue. Tout dépendra alors du degré de réactivité des communautés linguistiques et des États pour s'emparer de ces innovations et les traduire en une synergie multilingue réelle sur les réseaux.

Pour les moins concernés par les questions techniques et encore moins par la diversité linguistique numérique, le sujet des IDN est une problématique secondaire, voire inappropriée et parfois incompréhensible, étant donné que les interfaces homme-machine, les contenus d'information et les moteurs de recherche sont largement multilingues depuis les années 2000. Ils donnent ainsi la preuve que beaucoup ont encore tendance à oublier que tout le monde n'est pas censé connaître l'alphabet latin et qu'une partie importante de la population mondiale n'est pas alphabétisée. Pour les uns et les autres, des solutions techniques devraient pourtant être trouvées pour les impliquer dans la société du savoir partagé sachant bien que la possibilité d'utiliser sa langue maternelle sur les réseaux mondiaux d'information est de plus en plus considérée comme un indicateur de développement économique et culturel. L'Unesco en a fait un évènement annuel depuis 2000 sous forme d'une « Journée internationale de la langue maternelle ». Dans son discours prononcé à l'occasion de l'ouverture de la Journée en février 2011, sa Directrice générale souligne l'importance du maintien de cette richesse culturelle et linguistique : « chaque langue est unique dans sa façon de comprendre, d'écrire et d'exprimer la réalité ». Chaque langue maternelle est aussi un vecteur principal de transmission du savoir et des traditions. L'idéal serait donc d'harmoniser la langue d'une ressource avec la langue de son adressage sur Internet et celle de son utilisateur.

---

[14] BORTZMEYER (Stéphane), « La gouvernance de l'internet à l'heure du multilinguisme », in VANNINI (Laurent) ; LE CROSNIER (Hervé), Éd. *Net.Lang, op. cit.*, p. 410.

**Les promesses de l'audiovisuel numérique pour les langues orales en danger**

Unicode contribuera certainement à préserver beaucoup de langues de l'oubli en introduisant leurs modes d'écriture sur les réseaux et dans les systèmes informatiques. Mais il ne détiendra pas à lui seul la responsabilité de la présence de toutes les langues sur Internet car beaucoup d'entre elles sont restées exclusivement orales et donc inadaptées à toute forme d'encodage numérique textuel. Par conséquent, le multilinguisme numérique ne peut être confiné à la seule conception d'une saisie d'un contenu textuel à travers un clavier. De nouveaux artefacts autres que textuels émergent constamment dans les vagues récurrentes de l'innovation technologique comme la saisie et la lecture par analyse vocale, la voix et l'imagerie de synthèse, les commandes vocales, le rayon laser, les ondes radio…. Ces artefacts sont de plus en plus intégrés dans les outils numériques modernes comme les téléphones portables, les tablettes et les ordinateurs personnels. Certaines communautés de linguistes et de chercheurs rassemblant des corpus oraux, savent bien qu'à partir du moment où les enregistrements sonores sont désormais numériques et non plus analogiques, leur traitement est fondamentalement intégré dans un *continuum* multimédia numérique dans lequel le *distinguo* image, son, balisage, métadonnée, texte et paratexte n'a plus cours. Ces outils permettent d'étudier de façon fine et automatique le fil d'un discours, d'y repérer et documenter les pauses, les évènements, les caractéristiques, la translittération, la traduction, etc. Ils constituent aujourd'hui une méthode d'évidence plus économique de « faire de l'ethnolinguistique ».

Avec le potentiel du multimédia et de l'audiovisuel numérique, la multiplication des réseaux de transmission de données à haut débit, la radio et la télévision numériques, la téléphonie mobile puis un important arsenal normatif audiovisuel comme MPEG et MP3, les langues orales ont une nouvelle opportunité d'entamer un nouveau chapitre de leur lutte pour l'existence et la résistance à l'oubli. Avec chaque nouvelle innovation multimédia, l'espoir de voir les langues orales se doter d'une instrumentalisation plus efficace augmente.

Les voies d'expression et de communication sur Internet peuvent désormais prendre de la distance par rapport au clavier et au texte, prendre des formes audiovisuelles multiples par l'audio et la vidéo. Des services de communication comme la voix sur IP, la visiophonie par *Skype*, le *storytelling* ou la télévision et la radio numériques ainsi que les produits d'information audiovisuels du type *YouTube* et *Dailymotion* sont désormais d'un accès de plus en plus facile. Dès qu'une voix (discours véhiculant un message dans une langue) est enregistrée et transmise en mode analogique (radio) ou numérique (Internet), elle est relativement « sauvée ». Ce qui ferait la différence majeure pour optimiser sa reconnaissance et sa visibilité sont les conditions de son usage et de son exploitation.

Sur ce sujet, John Paolillo[15] est de l'avis que l'architecture même d'Internet, dans ses formats de documents, ses modes de transaction de contenus, ses artefacts d'entrée et de sortie des données, etc. peut ne pas toujours correspondre aux repères culturels et cognitifs des locuteurs d'une langue ou d'une culture donnée. Au-delà du rejet volontaire d'Internet au sein même des sociétés de culture latine[16], des cultures non occidentales enracinées dans la tradition orale et le contact direct peuvent refuser de s'inscrire dans la communication virtuelle et l'usage du multimédia et de l'audiovisuel numérique. Elles ressentiraient Internet comme incompatible avec leurs modes de vie et leurs valeurs traditionnelles. Paolillo, rapporté par Prado, cite l'exemple des Maoris qui, pour des questions purement culturelles, n'ont pas accepté les bibliothèques numériques. À partir de ce constat, « nous pouvons nous poser la question de savoir si la Toile, les forums, les blogues, les listes de diffusion ne vont pas parfois à l'encontre des principes ou valeurs culturelles d'un peuple, et de ce fait, seront moins utilisés par ces cultures, voire pas du tout […] Les formats propres à Internet sont loin d'être adaptés aux langues non écrites »[17].

---

[15] PAOLILLO (John), « Diversité linguistique sur Internet : examen des biais linguistiques », in *Mesurer la diversité linguistique sur Internet*, Paris, Unesco, 2005, p. 43.

[16] BOUDOKHANE (Feirouz), *L'Internet refusé : le non-usage du réseau et ses raisons*, Thèse, Université de Bordeaux 3, 2008

[17] PRADO (Daniel), *op. cit.*, p. 48

Abstraction faite de certains cas de résistance et de refus culturel, les produits et services multimédias qui sont proposés en permanence par les différentes vagues de l'innovation technologique constituent des alternatives prometteuses pour que des langues et des cultures orales, jusque-là inconnues, soient convenablement traitées et rendues visibles sur Internet. L'enregistrement sonore est souvent d'une meilleure authenticité linguistique et culturelle qu'une transcription écrite. Il trace dans ses plus petits détails la structure tonale et phonétique d'une langue, véhicule mieux qu'un texte écrit des données émotionnelles comme les intonations, les humeurs, les hésitations… et des données de contexte ambiant comme les bruits de fond. L'apport important que des techniques et des outils de traitement numérique d'enregistrements de signaux sonores et des corpus oraux, et aussi les perspectives de normalisation de ces outils aujourd'hui encore disparates, laisse envisager un retour en force de l'oral numérique.

En définitive, même si l'on sait bien que la numérisation d'une langue demeure fondamentalement une question technique (et industrielle), celle-ci ne se limite pas uniquement à un processus d'encodage binaire d'une écriture ou d'un enregistrement sonore. Elle suppose dans son ensemble – pour des langues en danger surtout – une série de mesures plus complexes et d'une stratégie d'action globale autour d'une politique d'aménagement linguistique numérique.

**L'aménagement linguistique : une géostratégie contre l'oubli des langues**

À notre sens, la vraie question de la vie ou de la mort des langues sur Internet ne peut être traitée en dehors de la question fondamentale de l'aménagement linguistique par lequel une écologie des langues s'opère dans une évolution « darwinienne » vers une modernité numérique vitale. Dans cette évolution, des langues, des dialectes ou des parlers sont condamnés à mourir par défaillance et inaptitude, mais beaucoup donnent aussi naissance à d'autres langues mieux adaptées et mieux préparées à une forme de modernité salvatrice. Une langue morte comme le latin, par exemple, a su donner naissance à beaucoup de langues européennes qui ont pu préserver un patrimoine commun.

Plusieurs langues actuelles sont dans la même situation. Elles seront certainement condamnées à mourir, mais le mérite qu'elles pourront avoir est de donner naissance à d'autres langues plus adaptées et mieux articulées avec leur temps et leur environnement.

La mort de langues et de systèmes d'écriture est plus fréquemment un problème de grandes zones linguistiques où ces langues et ces écritures n'ont pas pu développer une dynamique de survie. Des dialectes et des parlers inconnus ou ignorés voulant éviter l'exclusion numérique n'ont d'autres solutions que de se constituer en communautés linguistiquement rapprochées. Si une langue comme le berbère a pu être intégrée dans Unicode, alors qu'elle n'y était pas à l'origine du standard en raison de la variabilité de ses modes de transcription, c'est sans doute parce que les communautés berbérophones se sont entendues sur une seule transcription qui représenterait tous les parlers berbères dans l'univers numérique. Il est clair que les Berbères ont pris le problème à bras le corps pour se dire qu'on préfère considérer qu'il faut un mouvement centripète pour un petit nombre de grandes langues ou zones (kabyle, marocain moyen…) en aménagement ; mais que cette seule langue en aménagement assure la survie d'un patrimoine et d'une littérature orale ancienne justement car une langue aménagée permet de les transcrire. Depuis son intégration dans Unicode, des sites berbérophones se multiplient donnant à une langue une visibilité qu'elle n'avait pas il y a peu.

Cet exemple confirme que là où un aménagement linguistique numérique peut opérer, c'est là où la diversité n'a pas empêché le rapprochement de différentes langues et des dialectes même si celles-ci ont des disparités non seulement de vocabulaires, mais aussi de grammaire et de syntaxe. En créant un parler moyen et convergent, les langues se protègent contre l'isolement et parviennent à créer une réalité de communication et d'échange sur des aires qui sont raisonnables par rapport aux échanges médiatiques utilisés par la moyenne des gens.

L'aménagement linguistique n'est pourtant pas uniquement l'aménagement d'une transcription. L'*Encyclopédie Universalis* le définit comme un aménagement du corpus des langues « du point de vue graphique (création d'une écriture et d'un système de transcription, changement d'un type d'écriture à un autre, etc.), du point de vue orthographique (fixation de l'orthographe, création de grammaires, mise à jour, etc.) et du point de vue du lexique (création de lexiques, dictionnaires, terminologies spécialisées et de langue générale, harmonisation toponymique, etc.). L'aménagement de la circulation des langues dans les circuits de l'édition, des radios et télévisions, publicité, signalétique, etc. »[18].

Dans l'univers numérique, ce n'est qu'une continuation. La modernité numérique devrait permettre à des locuteurs de langues cousines de s'associer sur des langues de vastes espaces et de les standardiser pour arriver à communiquer. Aujourd'hui, l'espace Internet confronte perpétuellement les gens à communiquer sur un espace plus vaste via les vidéos numériques et les médias sociaux comme *Facebook*, *Skype* et *YouTube* alors qu'autrefois, c'était seulement le cinéma et la télévision.

La dynamique de ce nouvel aménagement linguistique numérique fera sans doute mourir des langues figées ou isolées mais participera à l'émergence de langues mieux adaptées à leur époque. Il s'agira d'une mort « bénéfique » de parlers ou de langues cousines pour aménager des langues moins nombreuses mais réellement vivantes. C'est le destin normal de l'adaptation (et de l'évolution darwinienne) des langues mères (disparates et plurielles dans leur cousinage) qui s'aménagent de façon volontaire ou implicite afin de répondre aux nouveaux défis de la modernité numérique.

## BIBLIOGRAPHIE

---

[18] DEPECKER (Loïc), « Aménagement linguistique », in *Encyclopædia Universalis*, [en ligne], s.d., p. 8, [http://www.universalis.fr/encyclopedie/amenagement-linguistique/]